\documentclass[11pt]{article}
\usepackage{amsmath, amsthm, amssymb}
\newcommand{\ket}[1]{\left | #1 \right \rangle}
\newcommand{\bra}[1]{\left \langle #1 \right |}

\newcommand{\st}[1]{ {\rm Stab}( #1 ) }

\def\openone{\leavevmode\hbox{\small1\kern-3.8pt\normalsize1}}

\def\cn{{\cal N}}
\def\cm{{\cal M}}

\def\cc{{\cal C}}
\def\cp{{\cal P}}

\def\ZZ{\mathbb{Z}}

\def\NN{\mathbb{N}}
\def\CC{\mathbb{C}}

\newtheorem{theorem}{Theorem}
\newtheorem{definition}{Definition}
\newtheorem{lemma}{Lemma}

\newtheorem{proposition}{Proposition}
\newtheorem{corollary}{Corollary}

\newcommand{\proj}[1]{\ket{#1}\!\bra{#1}}

\newcommand{\beq}{\begin{equation}}
\newcommand{\eeq}{\end{equation}}
\newcommand{\beqa}{\begin{eqnarray}}
\newcommand{\eeqa}{\end{eqnarray}}

\newcommand{\N}{\mathbb{N}}
\newcommand{\phases}{\mathbb{S}_1}
\newcommand{\gset}[1]{\left\langle #1 \right\rangle}
\newcommand{\set}[1]{\left\{#1\right\}}
\newcommand{\nzer}[2]{\mathcal{N}_{#1}(#2)}  % To symbolise the normaliser
\newcommand{\nzeri}[1]{\mathcal{N}(#1)}  % To symbolise the normaliser with implicit ambient group
\newcommand{\projnzeri}[1]{\mathcal{PN}(#1)}  % To symbolise the normaliser with implicit ambient group
\newcommand{\G}{\mathcal{G}}  % To symbolise mathcal G - for groups
\newcommand{\gate}[1]{\left( \begin{matrix} #1 \end{matrix} \right)}
\newcommand{\smallGate}[1]{\left( \begin{smallmatrix} #1 \end{smallmatrix} \right)}

\newcommand{\be}{\begin{equation}}
\newcommand{\ee}{\end{equation}}
\newcommand{\bea}{\begin{eqnarray}}
\newcommand{\eea}{\end{eqnarray}}
\newcommand{\bes}{\begin{equation*}}
\newcommand{\ees}{\end{equation*}}
\newcommand{\beas}{\begin{eqnarray*}}
\newcommand{\eeas}{\end{eqnarray*}}
\begin{document}
\begin{center}
{\LARGE\bf Generalised Clifford groups and simulation\\ of
associated quantum circuits\\ }
\bigskip
{\normalsize Sean Clark$^\dagger$, Richard Jozsa$^\dagger$ and Noah Linden$^\S$}\\
\bigskip
{\small\it $^\dagger$Department of Computer Science,
University of Bristol,\\ Merchant Venturers Building, Bristol BS8 1UB U.K. \\

$^\S$Department of Mathematics, University of Bristol,\\
University Walk, Bristol BS8 1TW U.K.}
\\[4mm]
\date{today}
\end{center}

\begin{abstract}
Quantum computations that involve only Clifford operations are
classically simulable despite the fact that they generate highly
entangled states; this is the content of the Gottesman-Knill
theorem.  Here we isolate the ingredients of the theorem and
provide generalisations of some of them with the aim of
identifying new classes of simulable quantum computations. In the
usual construction, Clifford operations arise as projective
normalisers of the first and second tensor powers of the Pauli
group.  We consider replacing the Pauli group by an arbitrary
finite subgroup $G$ of $U(d)$. In particular we seek $G$ such that
$G\otimes G$ has an entangling normaliser. Via a generalisation of
the Gottesman-Knill theorem the resulting normalisers lead to
classes of quantum circuits that can be classically efficiently
simulated. For the qubit case $d = 2$ we exhaustively treat all
finite subgroups of $U(2)$ and find that the only ones (up to
unitary equivalence and trivial phase extensions) with entangling
normalisers are the groups generated by X and the $n^{\rm th}$
root of $Z$ for $n \in \N$.

\end{abstract}

\section{Introduction}\label{intro}
The identification of classes of quantum computations that can be
classically efficiently simulated is a basic tool for studying the
relationship between classical and quantum computational power.
One of the earliest results in this context is the Gottesman-Knill
(GK) theorem \cite{Gthesis,NC}. It initially arose in the study of
the stabiliser formalism for quantum error correcting codes and
has a rich variety of mathematical ingredients. In this paper we
isolate these ingredients and develop generalisations of some of
them with an aim of identifying further new classes of simulable
quantum computations. (Other directions of generalisation of the
GK theorem were considered in \cite{AG}).

The study of classical simulation is closely related to the
invention of novel mathematical formalisms for the representation
of quantum states and computational steps, and the consequent
calculation of measurement probabilities. Indeed relative to any
such mathematical formalism there will be a class of states with
``small'' descriptions whose updates can be efficiently calculated
within that representation. The most commonly used formalism
describes states in terms of their amplitudes in the computational
basis and then the property of entanglement leads generically to
exponentially large descriptions, inhibiting efficient classical
simulation. Consequently\cite{JL} we may claim that entanglement
is an important resource for quantum computational power in the
sense that if it is absent then the quantum computation may be
classically efficiently simulated. However the situation becomes
less clear when we realise that in alternative formalisms the
class of states with suitably small descriptions can include rich
varieties of entangled states leading to the efficient simulation
of classes of computations that involve much entanglement along
the way. In the development of the theory of quantum error
correction, highly entangled quantum code states appeared
increasingly ungainly in the standard amplitude description. The
stabiliser formalism was introduced \cite{Gthesis} to provide a
compact and manageable description leading to the notion of
general stabiliser states, having small descriptions in this
formalism. Subsequently one could ask for quantum operations that
preserve this class of states and hence arrive at the GK theorem,
loosely speaking, that associated classes of quantum circuits
(although developing entanglement) can be efficiently simulated.

To motivate our proposed generalisations we first outline the
ingredients of the GK theorem. Let $X$ and $Z$ denote the standard
qubit Pauli operations. The Pauli group is defined by $\cp =
\langle X,Z, iI \rangle$ (where the pointed brackets denote the
group generated by the enclosed elements). The Pauli group on $n$
qubits is defined as the $n$-fold tensor power $\cp_n=
\cp^{\otimes n}$ which is a finite group of size $|\cp_ n| =
O(4^n)$. If $\ket{\psi}$ is any $n$ qubit state define its
stabiliser as
\[ \st{\psi}= \{ g\in \cp_n : g\ket{\psi}=\ket{\psi} \} . \]
Clearly $\st{\psi}$ is always a subgroup of $\cp_n$ (albeit the
trivial subgroup for many $\ket{\psi}$'s). $\ket{\psi}$ is a
stabiliser state if it is uniquely characterised by its stabiliser
i.e. it is the only state left invariant by all $g\in \st{\psi}$.
Any subgroup of a finite group $G$ may be described by $O(\log
|G|)$ generating elements\cite{NC}, providing our small
description of stabiliser states (which can generally be highly
entangled): $\st{\psi}= \langle g_1,\ldots ,g_r \rangle$ with
$r=O(n)$.

Next note that for any $U$, $g\ket{\psi}=\ket{\psi}$ iff
$(UgU^\dagger) U\ket{\psi}=U\ket{\psi}$ so corresponding to the
application of a gate $U$ to $\ket{\psi}$, $\st{\psi}$ is updated
by conjugation of the generators by $U$. In order to guarantee
that $U\ket{\psi}$ is again a stabiliser state we require that
$UgU^\dagger \in \cp_n$ for all generators $g$. To guarantee this
condition independently of the choice of $\ket{\psi}$ we impose it
for every $g\in \cp_n$ introducing the so-called Clifford
operations: a Clifford operation on $n$ qubits is a unitary
$n$-qubit operator $C$ with the property that $CgC^\dagger \in
\cp_n$ for all $g\in \cp_n$. For each $n$ we obtain the Clifford
group $\cc_n$ defined by
\[ \cc_n = \{ C\in U(2^n):C\cp_n C^\dagger = \cp_n \} \]
i.e. $\cc_n$ is the (group-theoretic) normaliser of the group
$\cp_n$ (within the unitary group). Let $H$ denote the Hadamard
operation, let $P$ denote the $\pi /4$-phase gate  \[ P=\left(
 \begin{array}{cc} 1 & 0 \\ 0 & i \end{array} \right) \] and let
$CZ$ denote the 2-qubit controlled-$Z$ gate. Then we have the
following full explicit characterisation \cite{cliffs,Gthesis} of
Clifford operations.
\begin{lemma}\label{clifflemma} $\cc_1=\langle H,P \rangle$ and
$\cc_2=\langle \cc_1\otimes \cc_1, CZ \rangle$. For $n\geq 3$ an
$n$-qubit gate $U$ is in $\cc_n$ iff it can be expressed as a
circuit of gates from $\cc_1$ and $\cc_2$.
\end{lemma}

In terms of these structures we can give a precise statement of
the GK theorem:
\begin{theorem} (Gottesman-Knill theorem). Consider any
polynomial-time quantum computation of the following sort. The
starting state $\ket{\psi_0}$ is a stabiliser state and each
computational step is one of the following:\\ (a) a measurement on
a qubit in the $Z$ basis, or\\ (b) application of a gate from
$\cc_1$ or $\cc_2$ (not depending on measurement outcomes in (a)), or\\
(c)  application of a gate from $\cc_1$ or $\cc_2$ chosen
adaptively depending on a previous measurement outcome from (a).\\
Finally the output is the result of a measurement of the first
qubit in the $Z$ basis.\\ Then the computation may be classically
efficiently simulated.
\end{theorem}

The standard proof of this result (see e.g. \cite{NC} or \cite{AG}
for a more recent improved algorithm) proceeds by updating the
stabiliser description of the state through the course of the
computation. The update procedure for (b) (and (c) once the
measurement result is given) is via the normalising property of
the Clifford group in relation to the Pauli group $\cp_n$
containing the generators. This purely group-theoretic
relationship in itself, may be entertained for any group $G$
replacing the Pauli group. On the other hand the stabiliser update
rules for (a) (as elaborated for example in \cite{NC} page 463)
depend on further features specific to the Pauli group, such as
the fact that in this group every two elements either commute of
anti-commute.

The starting point for our generalisations is an alternative
simpler proof in the absence of the adaptively chosen steps in
(c): instead of forwardly propagating the state description we
will backwardly propagate the final measurement allowing us in
particular even to free the simulation from requiring stabiliser
states. Thus let $C$ now be any circuit of Clifford operations on
starting state $\ket{\psi_0}$ which is now not required to be a
stabiliser state. If the final measurement on the first qubit has
outputs 0,1 with probabilities $p_0,p_1$ then $p_0-p_1$ is given
by the expectation value of $Z_1=Z\otimes I\otimes \ldots \otimes
I$ in the final state $C\ket{\psi_0}$:
\begin{equation}\label{probs} p_0-p_1 =\bra{\psi_0}C^\dagger Z_1
C\ket{\psi_0}. \end{equation} This computation suffices to
simulate the output (as we also have $p_0+p_1=1$). Now $Z_1\in
\cp_n$ so $C^\dagger Z_1C$ has the product form $P_{i_1}\otimes
\ldots \otimes P_{i_n}$ for Pauli operators $P_{i_k}$. Hence if
$\ket{\psi_0}$ is any product state $\ket{\psi_0}=\ket{a_1}\ldots
\ket{a_n}$ then we get
\begin{equation}\label{probcalc} p_0-p_1 =\prod_{k=1}^n \bra{a_k}
P_{i_k}\ket{a_k} \end{equation} which can clearly be calculated
classically in linear time $O(n)$. Similarly the commuting of the
successive one and two qubit Clifford gates through $Z_1$ can also
be done in time linear in the size of the circuit giving a linear
time classical simulation of the quantum computation's output.

This approach to the simulation of Clifford circuits may also be
extended to allow for measurement steps (of type (a) above) so
long as subsequent gates are not chosen adaptively (as they are in
(c) above, with stabiliser starting states). To achieve this we
replace each measurement step by the following: for each
measurement on a qubit $i$ adjoin an extra initial qubit in state
$\ket{0}$ and replace the measurement step by a (Clifford) CNOT
operation with control and target being the $i^{\rm th}$ and new
qubits respectively. The newly introduced qubit is not used in any
other way by the computation so its presence serves to decohere
the $i^{\rm th}$ qubit into the post-measurement mixture i.e. each
measurement step of the form (a) is replaced by a CNOT step of the
form (b) and the final output is unchanged. We may ask if a
further such trick could allow efficient simulation of the output
of the process with adaptively chosen gates (as in (c)) in
addition to just measurements (a) themselves, for the scenario of
Clifford circuits on arbitrary product starting states. Such
further generalisation is not likely to be possible for the
following reason: if we allow arbitrary product state inputs and
adaptive Clifford gate choices then we could (as shown in
\cite{BK}) implement the $\pi /8$-phase gate
\[ S=\left(
 \begin{array}{cc} 1 & 0 \\ 0 & e^{i\pi /4} \end{array} \right). \]

To see how this is achieved let
$\ket{\alpha}=\frac{1}{\sqrt{2}}(\ket{0}+e^{i\pi /4}\ket{1}$. Then
for any qubit $\ket{\psi}$ apply CNOT to $\ket{\psi}\ket{\alpha}$
and measure the second qubit. If the outcome is 0 then the
post-measurement state is $S\ket{\psi}\ket{0}$. If the outcome is
1 then the post-measurement state is
$S^\dagger\ket{\psi}\ket{1}e^{i\pi/4}$ so applying $P$ to the
first qubit gives $S\ket{\psi}\ket{1}$ up to an overall phase.
Thus we implement $S$ in either case by responding adaptively to
the measurement outcome. A supply of $\ket{\alpha}$ states can be
provided as an extension of the input product state. Now it is
known that $S$ together with $\cc_2$ is a universal set of gates
for quantum computation so we would get an efficient simulation of
all poly-time quantum computation, which is generally believed not
to be possible.

Our discussion above generalises the GK theorem by allowing
arbitrary product state inputs (and noting that also arbitrary
entangled stabiliser state inputs can be generated by a prefixed
Clifford circuit on the product state $\ket{0}\ldots \ket{0}$) but
on the other hand restricts the original form by not allowing
adaptive choices of gates. Its virtue is that it relies only on
the group-theoretic normaliser relationship between Clifford and
Pauli groups and may thus be immediately generalised to having
arbitrary unitary matrix groups $G$ replacing the Pauli group as
the starting point. We require no associated subgroup structure to
support a stabiliser state formalism nor any consideration of
stabiliser states themselves.

Let $G$ be any finite matrix subgroup of $U(d)$. Introduce the
(linear) normalisers of $G$ and $G\otimes G$:
\[ \begin{array}{c} \cn(G)=\{ U\in U(d): UGU^\dagger=G \} \\
\cn(G\otimes G)=\{ U\in U(d^2): U(G\otimes G)U^\dagger=G\otimes G
\} . \end{array}
\]
We will be interested in using normaliser operations as circuit
gates and we can therefore allow extra overall phases to be
generated in the above relations. Thus we introduce the notion of
{\em projective normaliser}:
\[ \begin{array}{c} \cp\cn(G)=\{ U\in U(d): \mbox{$\forall g\in
G$, $UgU^\dagger=cg'$ for some $g'\in G$ and $c\in \phases \} $ }\\
\cp\cn(G\otimes G)=\{ U\in U(d^2): \mbox{$\forall g\in G\otimes
G$, $UgU^\dagger=cg'$ for some $g'\in G\otimes G$ and $c\in
\phases \} $ }. \end{array}
\] where $\phases =\{ e^{i\theta}:0\leq\theta <2\pi \}$.\\
{\bf Remark}. The significance of projective normalisers is
illustrated by the following example. If $G$ is the Pauli group
$\langle X,Z,iI \rangle$ then $\cn (G)=\cp\cn (G)$ and it contains
the phase gate $P$. But if $G$ is the group $\langle X,Z \rangle$
comprising matrices with only real entries then $\cn (G)$ does not
contain the phase gate $P$ but $\cp\cn (G)$ does capture this gate
remedying the absence of complex elements in the centre of this
smaller real number version of the Pauli group. $\Box$

Mimicking our previous discussion we will especially seek examples
of groups $G$ such that $\cp\cn (G\otimes G)$ contains an
entangling gate (such as CNOT in the case of $G$ being the Pauli
group). Otherwise all normaliser circuits will preserve product
states and be computationally uninteresting. Furthermore in view
of eq. (\ref{probs}) it is desirable that $G$ contains a Hermitian
element (such as $Z$) which can be associated with a measurement.
Then we will be able to efficiently calculate its expectation
value in the final state of any normaliser circuit (with product
state input) forming the basis of our classical simulation
procedure. Even if $G$ does not contain a Hermitian element we may
use the Hermitian matrix $A+A^\dagger$ for any $A\in G$ and
similarly apply the arguments following eq. (\ref{probs}) to
simulate an associated measurement expectation value. In this vein
we also note that if $G$ acts irreducibly on $\CC^d$ (e.g. as is
the case for the usual Pauli qubit group) then any $d\times d$
matrix may be expressed as a linear combination of the matrices of
$G$ (c.f. \cite{serre} p. 48) so we may efficiently compute the
expectation value for {\em any} von Neumann measurement on a
single qudit or more generally on $O(\log n)$ qudits.

\subsection{Teleportation groups}\label{telepgroups}
In this paper we will consider only subgroups $G$ of $U(d)$ that
act {\it irreducibly} on $\CC^d$. In addition to facilitating the
mathematical analysis at various stages such groups have an extra
significance as prospective generalised substitutes for the Pauli
group as follows. Recall that another fundamental appearance of
the Pauli group is in quantum teleportation, providing the set of
Bob's ``correction operators''. In measurement based quantum
computation\cite{mmtcomp}, which can be viewed from the
perspective of teleportation \cite{mqc} the associated Clifford
operations have a special role of being parallelisable to depth 1
in this formalism. Thus we may ask what other sets of operators
may appear as Bob's correction operators in generalised
teleportation schemes and then ask for their normalisers. This
will again lead to classes of computations that are parallelisable
in the corresponding generalised measurement based computational
model. In this regard, irreducible subgroups of $U(d)$ play an
important role.

Let us define a generalised teleportation scheme as follows. Alice
and Bob share the maximally entangled 2-qudit state
$\ket{\phi}=\frac{1}{\sqrt{d}}\sum \ket{i}\ket{i}$ and Alice also
has a 1-qudit state $\ket{\alpha}$. Let $\cm = \{ A_1,\ldots ,A_r
\}$ be any 2-qudit generalised measurement (POVM). Suppose Alice
applies $\cm$ to the first two qudits of
$\ket{\alpha}_1\ket{\phi}_{23}$ obtaining measurement outcome
$A_i$. Let $\rho_i$ be Bob's post-measurement state. (We may
without loss of generality take the full post-measurement state to
be $\sqrt{A_i}\ket{\alpha}\ket{\phi}$ renormalised, and $\rho_i$
is obtained by tracing out the first two qudits). This comprises a
generalised teleportation scheme if there exists a family of
unitary operators $U_i$ parameterised by the measurement outcomes,
such that for all $\ket{\alpha}$ and all $i$, $\rho_i$ is the pure
state $U_i\ket{\alpha}$ i.e. $U_i^\dagger$ functions as Bob's
correction operator for measurement outcome $i$. In the case that
$\{ U_1,\ldots ,U_r \}$ also forms a group we have the following.

\begin{lemma} Let $G=\{ U_1,\ldots ,U_r \}$ be any finite subgroup
of $U(d)$ that acts irreducibly on $\CC^d$. Then there exists a
generalised teleportation scheme with $G$ comprising Bob's
correction operators.
\end{lemma}

\noindent {\bf Proof}: Define $\ket{a_i}=U_i^\dagger\otimes
I\ket{\phi}$ for $i=1,\ldots ,r$ and introduce the positive rank 1
operators $A_i=\frac{d^2}{|G|} \proj{a_i}$. Then using Schur's
lemma (by virtue of the irreducible action of $G$) we can see that
$\sum_i A_i=I_{d^2}$ so $\{ A_1,\ldots ,A_r \}$ is a (rank 1)
POVM. Furthermore a straightforward calculation gives \[
\sqrt{A_i}\ket{\alpha}_1\ket{\phi}_{23}=\frac{1}{\sqrt{|G|}}
\ket{a_i}_{12} U_i\ket{\alpha}_3 \] and thus Bob's
post-measurement state is $U_i \ket{\alpha}$ as required. Also
each measurement outcome occurs with equal probability $1/|G|$.
$\Box$

In view of this result we introduce the term {\em teleportation
group} to refer to any finite subgroup of $U(d)$ that acts {\em
irreducibly} on $\CC^d$.

\subsection{Outline of the paper}\label{paperoutline}
Returning to our primary motivation of classical simulation we
would ideally wish to find all teleportation groups $G$ in $U(d)$,
compute the projective normalisers of $G$ and $G\otimes G$ seeking
especially the cases of $G$ such that $\cp\cn (G\otimes G)$
contains an entangling gate. We refer to such groups as {\em
entangling teleportation groups}.

In the case of the Pauli group the projective normalisers are
known explicitly analytically. However the derivation is lengthy
and rests on many properties special to the Pauli operators. We
are not able to similarly explicitly analytically characterise
projective normalisers for general teleportation groups (even for
$d=2$) and we resort to exhaustive methods using various computer
algebra packages. In the qubit case $d=2$ we will be able to treat
exhaustively all possible teleportation groups. In section
\ref{projnorm} we will describe our algorithm for computing
normalisers and projective normalisers of $G$ and $G\otimes G$ for
any given teleportation group. In these methods it will be
important to cut down wherever conveniently possible, the range of
various cases that needs to be considered to allow the computer
algebra to terminate in a reasonable time. In this respect it is
important to note that the centre $Z(G)$ of any teleportation
group $G$ (which by Schur's lemma comprises only phase multiples
of the identity) plays no role in extending or limiting the
existence of projective normalisers. Hence in section
\ref{basegps} we will describe how a search over all teleportation
groups in $U(d)$ for entangling ones, can be reduced to the study
of normalisers of projectively inequivalent projective
representations of the so-called base groups in $U(d)$ which are
defined to be the central quotients of teleportation groups.

Next, in section \ref{ewetwo} we will apply our methods to
identify all possible entangling teleportation groups in the qubit
case of subgroups of $U(2)$. We prove the following
result.\begin{theorem} \label{mainthm} The only finite subgroups
$G$ of $U(2)$ (up to unitary equivalence and trivial phase
extensions) such that $G\otimes G$ has an entangling projective
normaliser, are $\langle X,Z^{1/n} \rangle $ for $n\in \NN $ (the
usual Pauli group being a central extension of the case $n=1$).
\end{theorem}

Finally in section \ref{conclusions} we will make some concluding
remarks and identify some avenues for further developments.

\section{Algorithm for determining normalisers and projective
normalisers}\label{projnorm}

In this section we describe a procedure for computing the linear
and projective normaliser elements of a teleportation group $G =
\set{U_j} \subset U(d)$

Let $Gen(G) \subset G$ be a set of generators for $G$. For each $U
\in Gen(G)$ let $U^\prime \in G$ denote the image of $U$ under
conjugation with some $N \in \nzeri{\G}$:
\be\label{eqn_general_nzer} NUN^\dagger = U^\prime. \ee Let us
rewrite the normaliser matrix $N$ as a  $(d^2 \times 1)$-column
vector $\vec{n}$ where \be\label{eqn_notation_n} \vec{n}_{dj + k}
= N_{j,k}. \ee Then eq. (\ref{eqn_general_nzer}) becomes
\be\label{eqn_vec_nzer} (I \otimes U^{\top} - U^\prime \otimes I)
\vec{n} = 0 \ee where $I$ is the $(d \times d)$ identity matrix.

By specifying the values of one such pair $U$ and $U^\prime$ and
treating the entries in the vector $\vec{n}$ as unknowns we can
obtain from eq. (\ref{eqn_vec_nzer}) $d^2$ simultaneous equations
in $d^2$ unknowns.  If we assign members of $G$ as the images of
all the elements of $Gen(G)$ then we can solve these equations
simultaneously by finding the null space of $(I \otimes U^{\top} -
U^\prime \otimes I)$ for each $U \in Gen(G)$ and its chosen image
$U^\prime$.  If a non=trivial solution exists for $\vec{n}$
simultaneously for all $U$ then this gives us a solution for $N
\in \nzeri{\G}$.

This provides us with the basis for an algorithmic approach to
computing the elements of the normaliser.  We enumerate all the
possible choices of images of the generators of $G$ and solve the
simultaneous equations, discarding the trivial solutions.  In
order to improve the performance of this approach we observe that
all the mappings on $G$ induced by conjugation with a normaliser
element $N$ are constrained by the fact that each image must have
the same order as the generator and any pairwise choice of images
must preserve the group relations of the corresponding generators.

Thus we get the following algorithm for computing the normaliser
elements of a teleportation group.

\textbf{Procedure 1 - Compute $\nzeri{G}$}
\begin{enumerate}
    \item For each $U$ in $G$ compute $Order(U)$, the elements of
    $G$ with the same order as $U$, $Comm(U)$, the elements of $G$
    that commute with $U$ and $NComm(U)$, the elements of $G$ that
    do not commute with $U$.
    \item Take a minimal set of generators $Gen(G) = \set{U_1, \dots
    U_r}$ of $G$ and find the set of all pairs that commute.
    \item Calculate all possible images of $Gen(G)$ by considering
    all the choices given in steps 4,5 and 6.
    \item For the images of the first generator, $U_1$, iterate
    through the set $Order(U_1)$.
    \item The set of possible images of each subsequent generator
    $U_j$ is formed by starting with the set $Order(U_j)$ and then
    repeatedly intersecting with $Comm(U_k)$ if $U_j$ and $U_k$
    commute and with $NComm(U_k)$ otherwise for each $k < j$.
    \item For each choice of possible images $\set{U_1^\prime,
    \dots U_r^\prime}$ of $\set{U_1, \dots U_r}$ compute the combined
    null space of $(I \otimes U_j^{\top} - U_j^\prime \otimes I)$
    for $j \in \set{1,\dots r}$.  Any non trivial solution corresponds
    to a normaliser element of $G$.
\end{enumerate}

We can apply the same procedure to compute $\cn (G\otimes G)$. In
addition we may also test to see if a normaliser gate is
entangling using the following result. A 2-qudit unitary operator
$V \in U(d^2)$ is said to be {\em entangling} if for all  $A, B
\in U(d)$ it is true that $V \neq A \otimes B$ and $V \neq SWAP(A
\otimes B)$ (where the $SWAP$ operation is defined by
$SWAP\ket{i}\ket{j}=\ket{j}\ket{i}$).  Then we have the following
characterisation \cite{universal_imprimitive}: $V$ is not
entangling if and only if one of the two following conditions
holds for every $i,j,k,l,\bar{i},\bar{j},\bar{k},\bar{l} \in
\set{0,\dots,d-1}$:
\begin{enumerate}
    \item $V_{ij,kl}V_{\bar{i}\bar{j},\bar{k}\bar{l}} = V_{i\bar{j},k\bar{l}}V_{\bar{i}j,\bar{k}l}$
    \item $V_{ij,kl}V_{\bar{i}\bar{j},\bar{k}\bar{l}} = V_{i\bar{j},\bar{k}l}V_{\bar{i}j,k\bar{l}}$
\end{enumerate}
By checking these simple algebraic conditions we can readily
identify if a given operation is entangling or not.

\subsection{Algorithm for projective normaliser
elements}\label{section_alg_phase} To develop an algorithm for
determining projective normalisers of a group $G$ we first show
that any such element can be found as a linear normaliser of a
group $G'$ generated by adding suitable additional central
elements to $G$.

Let $N$ be any (fixed, chosen) projective normaliser element for
$G$. Then for all $U\in G$
\begin{equation}\label{projeq} NUN^\dagger=cV \hspace{5mm}
\mbox{with $c\in \phases$ and $V\in G$} \end{equation} Since any
$U\in G$ has $U^{|G|}=I$, $c$ must be a $|G|^{\rm th}$ root of
unity. Thus if $G'$ is the group obtained by including all such
roots into $G$ we see that any operator is a projective normaliser
of $G$ iff it is a linear normaliser of $G'$. In practice
(especially when treating larger groups such as $G\otimes G$) this
extension of $G$ to $G'$ becomes too large to be manageable for
subsequent application of exhaustive enumerations in procedure 1.
Thus we develop more refined restrictions on $c$ to further limit
its possible values.

Note first that there is ambiguity in the choice of $c$ and $V$ in
eq. (\ref{projeq}) due to central phases that may already exist in
$G$. This is remedied using the following lemma.

\begin{lemma}\label{centres} If $G$ is any teleportation group
then the centres of $G$ and $G\otimes G$ are both cyclic,
comprising phase multiples of $I$.\end{lemma}

\noindent {\bf Proof}. Since $G$ acts irreducibly Schur's lemma
guarantees that any central element is a multiple of $I$. Thus
$Z(G)$ is a finite subgroup of $\phases$ and hence is cyclic
(necessarily generated by its element $e^{i\theta}$ with least
positive $\theta$.) For $G\otimes G$ let $g_1\otimes g_2$ be any
central element. Then it commutes with $g\otimes g_2^{-1}$ for all
$g\in G$ so $g_1\in Z(G)$. Similarly $g_2\in Z(G)$ so $Z(G\otimes
G)=Z(G)\otimes Z(G)$ is again a finite subgroup of $\phases$,
hence cyclic. $\Box$

Now let $\omega_sI$ with $\omega_s=e^{2\pi i/s}$ be the minimal
phase element of $Z(G)$. Then in eq. (\ref{projeq}) we choose
$c=e^{i\theta}$ such that $0\leq \theta <2\pi/s$ which fixes $c$
and $V$ uniquely. Furthermore with this choice, the unique
correspondence between $U$ and $V$ means that we can view $c$ as a
function of $V$ (rather than $U$):
\begin{equation}\label{uniqproj} NUN^\dagger = f(V)V\hspace{5mm}
\mbox{with $0\leq arg(f(V))<2\pi/s$.} \end{equation} (The function
$f$ will also depend on the choice of $N$ but we omit explicit
inclusion of this parameter for notational clarity.) We will also
refer to the association of phase values $f(V)$ to $V\in G$ as a
{\em phase function} for $G$. Introduce
\[ \Gamma = \{ f(V)V:V\in G \} = NGN^\dagger . \]
Thus $\Gamma$ is a unitary matrix group isomorphic to $G$ and
$Z(\Gamma)=Z(G)$.

\begin{lemma}\label{phases} Let $\{ U_1,\ldots ,U_r \}$ be any set
of generators for $G$.\\
(a) Then $\{ \omega_s,f(U_1)U_1,\ldots ,f(U_r)U_r \}$ generates
$\Gamma$.\\
(b) If $U_i$ has order $n_i$ then $f(U_i)$ has the form
$\omega_{sn_i}^k$ for some $0\leq k<sn_i$. \end{lemma}

\noindent {\bf Proof}. (a) For any $V\in G$ we have
$V=U_{i_1}U_{i_2}\ldots U_{i_m}=\prod_k U_{i_k}$. Also from eq.
(\ref{uniqproj}) $f:G\rightarrow \phases$ has the multiplicative
property: $f(V_1V_2)=f(V_1)f(V_2)z$ for some $z\in Z(G)$. Hence
$f(V)V=z\prod_k f(U_{i_k})U_{i_k}$ for some $z\in Z(G)$. Thus
$\omega_sI$ together
with $f(U_i)U_i$ for $i=1,\ldots ,r$ generates $\Gamma$.\\
(b) We have $f(U_i)^{n_i}U_i^{n_i}= f(U_i)^{n_i}I$ which is thus
in $Z(\Gamma)=Z(G)$. Hence $f(U_i)^{n_i}=\omega_s^k$ so $f(U_i)$
is a power of $\omega_{n_is}$. $\Box$

For any projective normaliser $N$ of $G$, lemma \ref{phases}
provides restrictions on the values that the phases $f(V)$ in eq.
(\ref{uniqproj}) can possibly take. Define $\Phi(G)$ to be the set
of all choices of $f(U_1),\ldots ,f(U_r)$ satisfying the
conditions (a) and (b) of the lemma. Extending $G$ by new central
elements $\Phi(G)$ will then give a group $G'$ whose linear
normalisers are precisely the projective normalisers of $G$. In
many practical examples $|Z(G)|=s$ and the generator orders $n_i$
are small compared to $|G|$ so the resulting extension to $G'$ can
be far smaller than that obtained by simply adding all $|G|^{\rm
th}$ roots of unity to $G$. Correspondingly we introduce the
following computational procedures.

\textbf{Procedure 2 - Compute $\Phi(G)$}
\begin{enumerate}
    \item Take a set of generators $\set{U_1, \dots, U_r}$ of $G$.
    Let the orders of the generators be $\set{n_1, \dots, n_r}$.
    \item Take a generating element $z$ of the centre of $G$.
    \item For each possible combination of $j_t \in \set{0 , \dots, n_ts-1}$
    for $t = 1, \dots,r$ perform steps 4 and 5.
    \item Let $f(U_t) = \omega_{n_ts}^{j_t}$.
    \item If $\gset{z,f(U_t)U_t: t \in \set{1,\dots,r}} \cong G$
    then add each $f(U_t)$ to the set $\Phi(G)$.
    \item Output $\Phi(G)$.
\end{enumerate}

We note that, for small groups, step 5 can be performed relatively
quickly using a computational package such as GAP\cite{gap}.

This provides us with an algorithm that produces a group
$G^\prime$ such that the projective normalisers of $G$ are the
linear normalisers of $G^\prime$ and hence a means to compute the
projective normaliser elements of a teleportation group and its
tensor square.

\textbf{Procedure 3 - Compute $\projnzeri{G}$}
\begin{enumerate}
    \item Compute $\Phi(G)$ using procedure 2.
    \item Compute $G^\prime = \gset{\phi I, U : \phi \in
    \Phi(G)\mbox{, } U \in G}$.
    \item Compute and output the linear normaliser elements
    of $G^\prime$ using procedure 1.
\end{enumerate}

Procedure 3 can be used to compute if teleportation groups are
projectively entangling in the following manner.  We compute the
two-qudit matrix group $G\otimes G$ from $G$ and use procedure 3
applied to the group $G\otimes G$ to find $\projnzeri{G\otimes G}$
(recalling that the centre of $G\otimes G$ is cyclic). Each
projective normaliser element found can then be tested to see if
it is entangling.

\section{Base groups and their projective
representations}\label{basegps}

Our principal aim is to apply the procedures of the preceding
section to systematically study classes of teleportation groups.
In the next section we will exhaustively treat the qubit case of
all teleportation subgroups of $U(2)$. There are infinitely many
teleportation groups in $U(d)$ but (at least for small $d$) they
are known to fall into regular families. Instead of directly
enumerating these we will adopt a different approach with a view
to reducing the amount of computer algebra required. It is clear
from the definition of $\cp\cn (G)$ that the group's centre $Z(G)$
plays no role in restricting or enabling new projective
normalisers.

For any teleportation group $G$ we introduce the central quotient
$B=G/Z(G)$. These central quotients are called {\em base groups}
of $U(d)$. (Our term ``base group'' is synonymous with ``finite
collineation group'' in \cite{blichfeldt}). Let $T:B\rightarrow G$
be any chosen transversal of $Z(G)$ in $G$ i.e. a choice of
element in each coset of $Z(G)$ in $G$. We also require that $I$
is chosen from $Z(G)$ itself. By slight abuse of notation we will
also use $T$ to denote the set of matrices $\{ T(b): b\in B \}$.
 Thus $T$ defines a projective representation of $B$ i.e. for
all $b_1,b_2\in B$ $T(b_1)T(b_2)=cT(b_1b_2)$ for some $c\in
\phases$. Conversely any projective representation $\rho$ of $B$
arises as a transversal of a representation of some $G$ with
$B=G/Z(G)$. (Indeed $G$ may be generated by the matrices of $\rho$
together with extra central phases $c$ from $\rho
(b_1)\rho(b_2)=c\rho (b_1b_2)$).

\begin{lemma} \label{gtob} Let $T$ be any projective
representation of $B=G/Z(G)$. Then $N$ is a projective normaliser
of $G$ (resp. $G\otimes G$) iff $N$ is a projective normaliser of
the set $T$ (resp. $T\otimes T= \{ A\otimes B: A,B\in T \}$).
\end{lemma}

\noindent {\bf Proof}: immediate from the fact that every $U\in G$
has the form $U=cV$ for some $V\in T$ and $c\in \phases$ (and
similarly for $G\otimes G$ and $T\otimes T$). $\Box$

Let $G_1$ and $G_2$ be teleportation groups with isomorphic
central quotients $B$. Choose transversals $T_1$ and $T_2$ giving
projective representations of the central quotients. We say that
$G_1$ and $G_2$ are {\em projectively equivalent} if the
projective representations $T_1, T_2$ (for some hence any choice
of transversals) are projectively equivalent as projective
representations of $B$ i.e. there is $A\in U(d)$ and $c(b)\in
\phases$ such that $T_1(b)=c(b)AT_2(b)A^\dagger$ for all $b\in B$.

Recall that $G$ is called entangling if $\cp\cn (G\otimes G)$
contains an entangling operation.

\begin{lemma} \label{preqgps} Let $G_1,G_2$ be projectively
equivalent teleportation groups. Then $G_1$ is entangling iff
$G_2$ is entangling. \end{lemma}

\noindent {\bf Proof}: Let $T_1,T_2,A$ be as above. Suppose $N$ is
an entangling projective normaliser for $G_1$. Introduce
$M=(A\otimes A) N (A^\dagger \otimes A^\dagger)$. Since $T_i$ is a
transversal of the centre of $G_i$ every member $U$ of $G_i$ has
the form $cV$ for $V\in T_i$ and $c\in \phases$. Thus a
straightforward calculation shows that $M$ is a projective
normaliser for $G_2\otimes G_2$. Also $M$ is locally equivalent to
$N$ so it is also entangling. $\Box$

In view of the above lemmas, to find all entangling teleportation
groups $G$ it suffices to look at a complete set of all
projectively inequivalent projective representations $T$ of all
base groups $B$ and determine if the projective normaliser of
$T\otimes T$ contains an entangling operation or not. $G$ up to
unitary equivalence is then generated by the matrices of $T$ and
further central phases $cI$. Actually even a complete list of
projectively inequivalent projective representations may involve
redundancies as the normaliser structure we seek is a property of
a {\em set} of matrices irrespective of how the set represents a
group. Then note that it is possible for two (projective)
representations of a group $B$ to be inequivalent yet comprise the
same overall set of matrices (up to overall phases) which are then
associated with the elements of $B$ in different ways (c.f eq.
(\ref{dihrep}) later for a non-projective example).

To access the full list of base groups for $d=2$ (and also for
$d=3,4$) we note that the full list of subgroups of $SU(2), SU(3),
SU(4)$ are known and provided in \cite{blichfeldt} (with the
latter two cases considered in more detail in \cite{subgroupsSU3}
and \cite{subgroupsSU4} respectively.)\\ {\bf Remark}. These lists
are complete up to abstract group isomorphism but not complete up
to unitary equivalence. In this regard it is important to note
that the normaliser group $\cn (G)$ is not a property of an
abstract group but of a given representation i.e. two unitarily
inequivalent representations of the same group will generally have
different
normalisers. $\Box$

To pass from base groups of $SU(d)$ to those of $U(d)$ we have the
following.

\begin{lemma}\label{sutou} The sets of central quotients of finite
subgroups of $SU(d)$ and $U(d)$ are the same. \end{lemma}

\noindent {\bf Proof}: If $G\subseteq SU(d)$ then $G\subseteq
U(d)$ so its central quotient is in both sets. Conversely if
$G\subseteq U(d)$ then the (finite) group $G'$ generated by
$det(g)^{-1/d}g$ for all $g\in G$ (and any choice of $d^{\rm th}$
root) has $G'\subseteq SU(d)$ and the same central quotient as $G$
i.e. $G/Z(G)$ also appears in both lists. $\Box$

Hence the full list of base groups of $U(2)$ (as abstract groups)
is obtained from the central quotients of the lists in
\cite{blichfeldt}.

In order to find all the projectively inequivalent irreducible
projective representations of a finite group we may use the
concept of a covering group \cite{curtisReiner, karpilovsky} and
it suffices to calculate the inequivalent irreducible linear
representations of this group.

\begin{definition} {(page 361 \cite{curtisReiner}) }
A covering group\footnote{Both `covering group'
\cite{curtisReiner} and `representation-group' \cite{karpilovsky}
are used in the literature as translations of Schur's
`Darstellungsgruppe'. We prefer `covering group' so as to avoid
the confusion of constructing a group representation of the
representation-group.} $G^{\star}$  of a finite group $G$ is a
finite group which is an extension of $G$ with kernel contained in
the centre of $G^{\star}$ such that every projective
representation of $G$ is equivalent to one which can be lifted to
a linear representation of $G^\star$.
\end{definition}
In this construction the linear representations of $G^*$ arise
from the projective representations of $G$ by inclusion of further
central elements (phase multiples of the identity).

The existence of such groups is then established by the following
theorem.
\begin{theorem}\label{thm_exist_covers}
Every finite group $G$ of order $n$ has at least one covering
group of order $nm$ where $m$ is the size of the Schur multiplier
(2nd cohomology group) of $G$.
\end{theorem}
\begin{proof}
Originally due to Schur. See Karpilovksy \cite{karpilovsky}.
\end{proof}

\begin{corollary}
In each dimension $d \ge 2$ there is a finite number of
projectively inequivalent irreducible unitary projective
representations of each finite group $G$.  Each of these can be
lifted to an irreducible unitary linear representation of a
covering group of $G$.
\end{corollary}
\begin{proof}
This follows directly from theorem \ref{thm_exist_covers}
specialised to irreducible unitary representations of a particular
dimension.
\end{proof}

Now we are in the position, for a given dimension $d$, to find all
the inequivalent teleportation groups using the following
procedure.

\textbf{Procedure 4 - Find projectively inequivalent teleportation
groups in $U(d)$}
\begin{enumerate}
    \item Let $S$ be a set of finite subgroups of $SU(d)$ up to isomorphism.
    \item Calculate all base groups of $U(d)$ as the set of central quotients of the elements of $S$.
    \item Let $B_C$ denote a set of covering groups for the base
    groups.
    \item A complete set of projectively inequivalent teleportation groups of $U(d)$ is
    generated by the set of inequivalent irreducible unitary linear representations of the
    elements of $B_C$.
\end{enumerate}

We now provide some additional practical information for the steps
in procedure 4.

\begin{enumerate}
    \item The lists of finite subgroups of $SU(d)$ for $d = 2,3,4$ are provided in the literature.
    \item In practice central quotients may be calculated using a computational package such as GAP \cite{gap}.
    The equivalence of the base groups of $SU(d)$ and $U(d)$ is shown in lemma \ref{sutou}.
    \item The existence of the covering groups in step 3 comes from theorem \ref{thm_exist_covers}
    and in practice we  use  GAP to calculate them.
    \item The set of inequivalent irreducible unitary representations of the covering groups can be produced in GAP.
    \item For each resulting representation $G$ we apply procedures 1,2,3 to
    determine whether $G\otimes G$ has an entangling projective normaliser
    or not.
\end{enumerate}

\subsection{Teleportation groups represented in
$GL(d,\CC )$}\label{section_gl} In applying the above procedures
we perform computations using the GAP computational system
\cite{gap} and in particular the REPSN \cite{repsn} representation
theory package to find the irreducible representations of finite
groups. For some groups REPSN does not produce a unitary
representation but a general linear representation in $GL(d,\CC
)$. In this section we relate the previous results concerning the
normaliser and projective normaliser of faithful irreducible
unitary representations to their general linear counterparts and
show that the algorithms we provide may be performed with faithful
irreducible representations of a finite group in $GL(d,\CC )$ and
the results applied to the faithful irreducible unitary
representations.

Two general linear representation $\rho^1_L, \rho^2_L:G\rightarrow
GL(d,\CC ) $ of a finite group $G$ are said to be equivalent if
there exists $M \in GL(d,\CC )$ such that for all $g \in G$ \be
\rho^1_L(g) = M \rho^2_L(g) M^{-1}. \ee

\begin{theorem}\label{thm_wigner_conj}
For any finite group $G$ and faithful irreducible representation
$\rho_L: G \rightarrow GL(d,\CC )$ of $G$ there exists a faithful
irreducible unitary representation $\rho_U:G \rightarrow U(d)$ and
a matrix $E \in GL(d,\CC )$ such that for all $g \in G$ \be
\rho_U(g) = E \rho_L(g) E^{-1}. \ee
\end{theorem}

\noindent {\bf Proof}: see \cite{wigner} page 74. $\Box$

Furthermore if $\rho_L$ in the theorem ranges over a full list of
inequivalent linear representations then $\rho_U$ ranges over a
full list of unitarily inequivalent representations. For general
linear representations we can define linear and projective
normalisers as follows.
\begin{equation}\label{linnorm}
\nzer{GL(d,\CC )}{\rho_L} = \set{N \in GL(d,\CC ) : N \rho_L(g)
N^{-1} \in \rho_L(G) \mbox{ for all } g \in G}. \end{equation}
\begin{equation}\label{projlinnorm}
\cp\nzer{GL(d,\CC )}{\rho_L} = \set{ N \in GL(d,\CC ) : \forall
g\in G \,\, \exists g'\in G, c\in \CC : N \rho_L(g) N^{-1}
=c\rho_L(g') }
\end{equation}
Theorem \ref{thm_wigner_conj} immediately gives:
\begin{corollary}
Given the two representations $\rho_L$ and $\rho_U$ of theorem
\ref{thm_wigner_conj} then every normaliser element $N \in
\nzer{U(d)}{\rho_U}$ defines a normaliser element $M = E^{-1}NE
\in \nzer{GL(d,\CC )}{\rho_L}$.
\end{corollary}

Note that in eq. (\ref{projlinnorm}) $c$ must actually be in
$\phases$ since $G$ is finite. Thus our previous procedures 1,2,3
can be used unchanged to compute the normalisers and projective
normalisers of $\rho_L$ and $\rho_L\otimes \rho_L$. If the latter
fails to be entangling we can conclude by the corollary that
$\rho_U$ is also not entangling, without having to carry out the
translation from $\rho_L$ to $\rho_U$ explicitly.

\section{Application to teleportation groups in
$U(2)$}\label{ewetwo} We now apply our procedures to consider all
teleportation groups in $U(2)$ and show that the entangling ones
are exactly those which have a central quotient isomorphic to a
dihedral group of order $4m$ for some integer $m$. Alternatively
these teleportation groups may be described as unitary equivalents
of groups obtained by adjoining additional central elements to the
matrix groups $\langle X, Z^{1/n} \rangle$ for $n\in \NN$.

We use a labelling system for small finite groups that is used in
 the GAP computational package.  This system assigns two numbers to
 a group.  The first is the order of the group and the second is a
 unique index for each group of a particular order.  As examples
 we have
     $[4,1]$ for the cyclic group of order 4,
     $[4,2]$ for the Klein four group,
      $[12,3]$ for the alternating group on 4 elements etc.

 \subsection{Base groups of $U(2)$}
 We use the list of base groups of $SU(2)$ which are given in
 Blichfeldt \cite{blichfeldt} as the finite collineation groups and
 which we are also the base groups of $U(2)$.  These consist of two
 infinite families of groups and three `special' groups.  The
 infinite families are the cyclic groups, which have no irreducible
 representations in $U(2)$, and the dihedral groups $D_{2n}$  of order
 $2n$.  The three special groups are the tetrahedral group $A_4
 \cong [12,3]$, the octahedral (or cube) group $S_4 \cong [24,12]$
 and the dodecahedral (or icosahedral) group $A_5 \cong [60,5]$.
 First we deal with the three special groups and finally the
 dihedral case.

 \subsection{The tetrahedral group as base group}
 A covering group of the tetrahedral group $A_4 \cong [12,3]$ is
 $[24,3]$ \cite{karpilovsky}.  We use the computational package GAP
 to compute three inequivalent representations in $U(2)$ which is
 the maximum number of inequivalent representations.  These are the
 matrix groups $M_1$, $M_2$ and $M_3$ where \be M_1 =
 \gset{\frac{1}{\sqrt{2}}\gate{\omega_8 & \omega_8 \\ \omega_8^3 &
 \omega_8^7}, \gate{-i & 0 \\ 0 & i}} \ee \be M_2 =
 \gset{\frac{1}{\sqrt{2}}\gate{\omega_{24}^{11} & \omega_{24}^{11}
 \\ \omega_{24}^{17} & \omega_{24}^5}, \gate{-i & 0 \\ 0 & i}} \ee
 \be M_3 = \gset{\frac{1}{\sqrt{2}}\gate{\omega_{24}^{19} &
 \omega_{24}^{19} \\ \omega_{24} & \omega_{24}^{13}}, \gate{-i & 0
 \\ 0 & i}}. \ee

 Using an implementation of the algorithm to find normalisers given
 in section \ref{projnorm} it can be seen that $M_1$, $M_2$
 and $M_3$ are not entangling.  To see that these three
 representations are not projectively entangling the ranges of the
 possible phase functions of the matrix groups $M_1$, $M_2$ and
 $M_3$ are calculated using the algorithm described in section
 \ref{section_alg_phase}.  From this we find that the possible
 phase functions take values in $\set{\omega_3^j: j=0,1,2}$.
   When we add the central elements corresponding to
 these phases to the matrix groups $M_1, M_2$ and $M_3$ we get
 matrix groups isomorphic to $Z_6 \otimes [24,3] \cong [72, 25]$.
 This group has only one irreducible representation in $U(2)$ up to
 equivalence.  This can be represented as \be
 \gset{\frac{1}{\sqrt{2}}\gate{\omega_{24}^{11} & \omega_{24}^{11}
 \\ \omega_{24}^{17} & \omega_{24}^5}, \gate{\omega_{12} & 0 \\ 0 &
 \omega_{12}^7}} \ee and we compute that it is not entangling.
 Hence no teleportation group in $U(2)$ with central quotient
 isomorphic to the tetrahedral group is entangling.

 \subsection{The octahedral group as base group}
 The octahedral group $S_4 \cong [24,12]$ has  $[48,29]$ as a
 covering group.   This has one representation in $U(2)$ up to
 equivalence which is not entangling.  The generators we used are
 \be \gset{ \frac{1}{\sqrt{2}} \gate{1 & -i \\ i & -1},
 \frac{1}{\sqrt{2}} \gate{\omega_{8}^{3} & \omega_8^7 \\ \omega_8^5
 & \omega_8^5}} \ee All possible phase functions take values in
 $\set{1,i}$ so to test to see if this group is projectively
 entangling we must test if any linear representation of $\ZZ_4 \otimes
 [48,29] \cong [96,192]$ is entangling.  Up to equivalence there
 is only one faithful representation in $U(2)$ and that has
 generators \be \gset{\frac{1}{\sqrt{2}}\gate{1 & -i \\ i & -1},
 \frac{1}{\sqrt{2}} \gate{\omega_{8} & \omega_{8}^{5} \\
 \omega_{8}^{3} & \omega_{8}^3}}. \ee We have computed that this
 matrix group is not entangling.  We conclude that no teleportation
 group with central quotient isomorphic to the octahedral group is
 entangling.

 \subsection{The dodecahedral group as base group}
 The dodecahedral group $A_5 \cong [60,5]$ has unique covering
 group $[120,5]$.  There are two faithful irreducible
 representations of this group up to equivalence in $U(2)$.  The
 computational system gap gives us representations in $GL(2,\CC )$
 which by the results of section \ref{section_gl} will suffice for our
 calculations.  The first representation has the following two
 generators \be
 \gate{\omega_{15}-\omega_{15}^2+\omega_{15}^4-\omega_{15}^8-\omega_{15}^{11}
 -\omega_{15}^{14} & -2\omega_{15}-2\omega_{15}^4-\omega_{15}^7-\omega_{15}^{13} \\
     -\omega_{15}^{11}-\omega_{15}^{14} & -\omega_{15}-\omega_{15}^4-\omega_{15}^7
     +\omega_{15}^{11}-\omega_{15}^{13}+\omega_{15}^{14} }
 \ee \be \gate{
 -\omega_{15}+\omega_{15}^2-\omega_{15}^4+\omega_{15}^8+\omega_{15}^{11}+\omega_{15}^{14}
 & \omega_{15}+\omega_{15}^4 \\
 \omega_{15}^2+\omega_{15}^8+2\omega_{15}^{11}+2\omega_{15}^{14} &
 \omega_{15}+\omega_{15}^4+\omega_{15}^7-\omega_{15}^{11}+\omega_{15}^{13}-\omega_{15}^{14}
 }. \ee The second representation has the following two generators
 \be \frac{1}{2}\gate{
  -2\omega_{15}-2\omega_{15}^4-\omega_{15}^7-\omega_{15}^{11}-\omega_{15}^{13}-\omega_{15}^{14}
   & 2\omega_{15}^2+\omega_{15}^7+2\omega_{15}^8+\omega_{15}^{11}+\omega_{15}^{13}+\omega_{15}^{14}  \\
     -\omega_{15}^7+\omega_{15}^{11}-\omega_{15}^{13}+\omega_{15}^{14} & \omega_{15}^7-\omega_{15}^{11}
     +\omega_{15}^{13}-\omega_{15}^{14}
 } \ee \be \frac{1}{2}\gate{
 \omega_{15}+\omega_{15}^2+\omega_{15}^4+\omega_{15}^7+\omega_{15}^8+2\omega_{15}^{11}
 +\omega_{15}^{13}+2\omega_{15}^{14}
 & \omega_{15}-\omega_{15}^2+\omega_{15}^4
  +\omega_{15}^7-\omega_{15}^8+\omega_{15}^{13} \\
     \omega_{15}-\omega_{15}^2+\omega_{15}^4+\omega_{15}^7-\omega_{15}^8+\omega_{15}^{13}
     & \omega_{15}-\omega_{15}^2+\omega_{15}^4-\omega_{15}^7-\omega_{15}^8
  -\omega_{15}^{13}
 }. \ee We have computed that these two representations are not
 entangling. Furthermore we have computed that all phase functions
 of both representations are trivial and so all two-qubit
 projective normaliser elements of are linear normaliser elements.
 This implies that the two representations of $[120,5]$ given above
 are not projectively entangling.  Hence no teleportation group in
 $U(2)$ which has central quotient isomorphic to the dodecahedral
 group is entangling.

 \subsection{A dihedral group as base group}

Since the family of dihedral groups comprises an infinite list we
approach the study of their projectively inequivalent projective
representations analytically.

 The Schur multiplier of a group \cite{karpilovsky} is key in
 calculating covering groups.  In particular when the Schur
 multiplier of a group is the trivial group then the group is its
 own covering group.  The Schur multiplier $M(D_{2n})$ of the
 dihedral group $D_{2n}$ is given in \cite{discreteTorsion} as \be
 M(D_{2n}) = \ZZ_{gcd(2,n)}. \ee This splits the analysis of the
 dihedral group $D_{2n}$ into the case of odd $n$ where $D_{2n}$
 covers itself and even $n$ where we obtain the so called binary
 dihedral groups as covering groups.

 \subsubsection{$D_{2n}$ when $n$ is odd}
 We claim that there are no entangling teleportation groups with
 central quotient isomorphic to $D_{2n}$ when $n$ is odd.

 As noted above, $D_{2n}$, for odd $n$, is
 its own covering group.  The only representations we need to
 consider in looking for entangling teleportation groups are the
 the irreducible linear ones of $D_{2n}$.  When $D_{2n}$ is
 presented as \be\label{eqn_present_di} D_{2n}= \gset{a,b | a^{n} =
 1, b^2 = 1, bab = a^{-1}} \ee we find, from \cite{serre}, the
 $r^{th}$ irreducible representation $\rho_r$ with $0 < r <
 \frac{(n-1)}{2}$ may be taken to be \begin{equation}\label{dihrep}
  \rho_r(a) = \gate{
 \omega_n^{r} & 0 \\ 0 & \omega_n^{-r}}\mbox{, }\rho_r(b) = \gate{
 0 & 1 \\ 1 & 0}. \end{equation}
  In particular we are only interested in the
 faithful representations when $gcd(r,n) = 1$.  We prove that the
 teleportation group $G$ for $r=1$ is not projectively entangling
 and the argument for general $r$ follows similarly. Indeed for
 each $r$ we get the same {\em set} of matrices (but associated to
 group elements in different ways) and the normalising property
  is a property of the collection of matrices only as a set.
 \begin{proposition}
 Let $G = \gset{A =\smallGate{ \omega_n & 0 \\ 0 & \omega_n^{-1}},
 B = \smallGate{ 0 & 1 \\ 1 & 0}} \cong D_{2n}$ for odd $n$.  Then
 every phase function of $G$ takes values in $\set{1,-1}$.
 \end{proposition}
 \begin{proof}
 For any phase function $f$ of $G$ we must have that \be f(A) =
 \omega_n^j \mbox{ and } f(B) = (-1)^k \ee by the orders of $A$ and
 $B$ respectively.  This defines the value of $f$ on all elements
 of $G$.  It is easily verified that if $j = 0$ and $k = 1$ then
 $f$ defines a valid phase function of $G$ but that for $j \ne 0$
 the group $\gset{f(A)A,f(B)B}$ is never isomorphic to $G$. $\Box$
 \end{proof}

 Since every phase function takes values in $\set{1,-1}$ we must
 now show that $G^\prime$ is not entangling where $G^\prime$ is
 generated by adding the element $\smallGate{-1 & 0 \\ 0 & -1}$ to
 the generators of $G$ .  So we have \be \label{gprime}
 G^\prime = \gset{C=
 \gate{\omega_{2n} & 0 \\ 0 & \omega_{2n}^{-1}}, B = \gate{ 0 & 1
 \\ 1 & 0}} \cong D_{4n}. \ee
 \begin{proposition}\label{prop_dihedral_odd}
 $G^\prime$ is not entangling for any odd $n \ge 3$.
 \end{proposition}
 The proof of the above is split into three lemmas from which the
 result follows.

 \begin{lemma}\label{lem_gen_perm}
 Every two-qubit normaliser gate for $G^\prime$ is a generalised
 permutation matrix.  That is it has exactly one non-zero entry for
 each row and each column.
 \end{lemma}
 \begin{proof}
 Let us write a general two-qubit normaliser gate of $G^\prime$ as
 $N$ with matrix entries $N_{jk}$.  We write $C_1 = C \otimes I$
 and $C_2 = I \otimes C$ with $C$ as in eq. (\ref{gprime}).
   For any $j \in \set{0,1,2,3}$ we have
 \be (N C_1 N^\dagger)_{jj} = \omega_{2n}( |N_{j1}|^2 + |N_{j2}|^2)
 + \omega_{2n}^{-1}( |N_{j3}|^2 + |N_{j4}|^2). \ee Since $N C_1
 N^\dagger \in G^\prime$ each entry $(N C_1 N^\dagger)_{j}$ must
be zero or a power of $\omega_{2n}$ and since $n \ge 3$ we must
have either \be\label{eqn_one_two}
 |N_{j1}|^2 + |N_{j2}|^2 = 0 \mbox{ or }|N_{j3}|^2 + |N_{j4}|^2 = 0.
\ee Similarly \be (N C_2 N^\dagger)_{jj} = \omega_{2n}( |N_{j1}|^2
+ |N_{j3}|^2) + \omega_{2n}^{-1}( |N_{j2}|^2 + |N_{j4}|^2) \ee
from which we conclude that either \be\label{eqn_one_three}
 |N_{j1}|^2 + |N_{j3}|^2 = 0 \mbox{ or }|N_{j2}|^2 + |N_{j4}|^2 = 0.
\ee From equations \ref{eqn_one_two} and \ref{eqn_one_three} we
see that $N$ must have exactly one non-zero entry per row and
since $N$ must be unitary we conclude it is a generalised
permutation matrix.$\Box$
\end{proof}
\begin{lemma}\label{lem_diag}
If there exists an entangling two-qubit normaliser gate for
$G^\prime$ then there also exists a diagonal entangling two-qubit
normaliser gate for $G^\prime$.
\end{lemma}
\begin{proof}
Suppose that $N$ is any entangling 2-qubit normaliser for $G'$.
From lemma \ref{lem_gen_perm} we have $N=DP$ where $D$ is diagonal
and $P$ is a permutation. The three 2-qubit normaliser gates
$I\otimes B$, $B\otimes I$ and $B\otimes B$ (with $B$ as in eq.
(\ref{gprime})) are permutations that interchange 00 with 01,10
and 11 respectively. Hence if $R$ is a suitably chosen one of
these three, then we get an entangling 2-qubit normaliser
$N'=NR=DP'$ where $P'$ is a permutation that leaves 00 fixed. If
$S$ denotes the swap gate (which is a 2-qubit normaliser for any
group) and $C_X$ denotes the controlled NOT gate, then the six
possible choices of $P'$ can be written as $I,S,C_X,SC_X,C_XS$ and
$SC_XS$. If $P'=I$ then $N'$ is diagonal. If $P'=S$ then $SN'$ is
a diagonal entangling normaliser. For $P'=C_X$ recall
$C_2=I\otimes C$ (with $C\in G'$ as in eq. (\ref{gprime})). Then a
direct calculation shows \begin{equation} N' C_2 N'^\dagger =
diag(\omega_{2n},\omega_{2n}^{-1},\omega_{2n}^{-1},\omega_{2n})
\notin G^\prime. \end{equation} Hence we cannot have $P'=C_X$.
Similarly if $P'$ were $SC_X,C_XS$ or $SC_XS$ we could pre- and/or
post-multiply $N'$ by $S$ to obtain a normaliser again of the form
$N''=D'C_X$ with $D'$ diagonal. Hence these three cases of $P'$
are also excluded and in all allowed cases, the existence of $N$
implies the existence of a diagonal 2-qubit entangling normaliser.
$\Box$
\end{proof}

\begin{lemma}\label{lem_diag_product}
No diagonal two-qubit normaliser gates for $G^\prime$ are
entangling.
\end{lemma}

\begin{proof}
Let us take an arbitrary diagonal two-qubit normaliser gate of
$G^\prime$ which we may write up to phase as $D = diag(1,a,b,c)$.
We then see that for $D B_1 D^\dagger$ to be in $G^\prime$ there
must be $j,k$ such that \be DB_1D^\dagger = \smallGate{ 0 & 0 &
\bar{b} & 0 \\ 0 & 0 & 0 & a\bar{c} \\ b & 0 & 0 & 0 \\ 0 &
c\bar{a} & 0 & 0} = C^jB\otimes C^k =  \smallGate{ 0
& 0 & \omega_{2n}^{j+k} & 0 \\ 0 & 0 & 0 & \omega_{2n}^{j-k} \\
\omega_{2n}^{-j+k} & 0 & 0 & 0 \\ 0 &\omega_{2n}^{-j-k} & 0 & 0}.
\ee Eliminating $b$ from the above we see that since $n$ is odd we
must have $k = 0$ or $k = n$.  Similarly from $DB_2D^\dagger$ we
have $l,m$ such that \be DB_2D^\dagger = \smallGate{ 0 & \bar{a} &
0 & 0 \\ a & 0 & 0 & 0 \\ 0 & 0 & 0 & b\bar{c} \\ 0 & 0 & c\bar{b}
& 0} = C^l \otimes C^mB =  \smallGate{ 0 & \omega_{2n}^{l+m} & 0 &
0 \\ \omega_{2n}^{l-m} & 0 & 0 & 0 \\ 0 & 0 & 0 &
\omega_{2n}^{-l+m} \\ 0 & 0 & \omega_{2n}^{-l-m} & 0} \ee and
eliminating $a$ gives $l = 0$ or $l = n$.  By solving for $a,b,c$
in the four cases of $k,l = 0,n$ we see that in each case $D$ can
be written as a tensor product of two gates and hence is not
entangling. $\Box$
\end{proof}

We have now completed the proof of proposition
\ref{prop_dihedral_odd}.  This completes the result that no
dihedral group $D_{2n}$ when $n$ is odd forms the base group of an
entangling teleportation group.

\subsubsection{$D_{2n}$ when $n$ is even}
We claim that every dihedral group $D_{2n}$, where $n$ is even, is
isomorphic to the central quotient of an entangling teleportation
group. Indeed introduce $m$ defined by $n=2m$ and
\[ Z^{1/m}=\gate{1 & 0 \\ 0 & \omega_{2m}} \hspace{5mm}
G_m=\langle X,Z^{1/m} \rangle .\] A straightforward calculation
shows that $\rho$ defined by $\rho(a)=Z^{1/m}$ and $\rho(b)=X$
provides a projective representation of $D_{2n}$ as presented in
eq. (\ref{eqn_present_di}) and $G_m/Z(G_m) \cong D_{2n}$.

For $m=1$ the projective normalisers are clearly just those of the
Pauli group given in lemma \ref{clifflemma}.

For $m\geq 2$ we find that $Z^{1/2m}= {\rm diag}(1,\omega_{4m})$
is a normaliser of $G_m$. This is the generalisation of the phase
gate $P$ from the Pauli group normaliser (and now $H$ is no longer
a normaliser). Also $CZ$ is a normaliser of $G_m\otimes G_m$ so
$G_m$ is entangling for all $m\in \NN$. Our computer algebra
procedures for small $m$ values showed that all normalisers of
$G_m\otimes G_m$ are generated from $CZ$ with SWAP and $G_m\otimes
G_m$ included.

Finally we show that there are no other teleportation groups
$G\subset U(2)$ with central quotient $D_{2n}$ ($n$ even) that are
not unitary equivalents of central extensions of $G_m$ above.
Indeed the covering group of $D_{2n}$ ($n$ even) is the binary
dihedral group $Q_{4n}$ with presentation\cite{MS}
\[ Q_{4n}= \{ a,b: a^{2n}=1, b^2=a^n, b^{-1}ab=a^{-1} \} \]
and a complete set of faithful irreducible representations on
$\CC^2$ is given by\cite{MS}
\[  \rho_r(a) = \gate{
 \omega_{2n}^{r} & 0 \\ 0 & \omega_{2n}^{-r}}\mbox{, }\rho_r(b) = \gate{
 0 & (-1)^r \\ 1 & 0}\hspace{2mm} \mbox{with gcd$(r,2n)=1$.} \]
 The case $r=1$ reproduces $G_{m}$ above and for each $r$,
 $\rho_r$ comprises the same set of matrices (up to phase
 multiples) so we get no new projective normaliser structures as
 $r$ varies.
This completes the proof of theorem \ref{mainthm}.

\section{Conclusions}\label{conclusions}

We have identified all entangling teleportation groups in $U(2)$
and seen that they comprise only a mild generalisation of the
standard qubit Pauli group. Also the associated projective
normalisers, apart from extra roots of $Z$, are already present
for the Pauli group case. Thus the qubit case appears to be of
limited scope in generating new classes of classically simulatable
circuits, but there are yet further cases and generalisations
worthy of investigation which we list as open questions.

Firstly we may consider higher values of $d$. The Pauli groups may
be naturally generalised to arbitrary dimension $d$ and the
associated normaliser groups were analytically characterised for
all prime $d$ in \cite{sean}. In this case it is found that an
analogue of lemma \ref{clifflemma} holds (with $H$, $P$ and $CZ$
being replaced by suitable one and two qudit operations as given
in \cite{sean}). We applied our computational programs to a
further few chosen examples of teleportation groups in $U(3)$ but
did not find any further interesting entangling ones. We were
unable to exhaustively treat all base groups of $U(3)$ because of
the increased size of the groups involved. Thus it would be
advantageous to further develop the study of computational
procedures for projective normalisers, inventing algorithms that
search over more restricted spaces of values.

We have considered the projective normaliser structure of $G$ and
$G\otimes G$ only for matrix groups $G$ that act irreducibly. But
more generally if a matrix group acts reducibly it is not clear
how its normaliser structure relates to that of its irreducible
parts. This may provide an avenue for constructing further
interesting examples of entangling normalisers for groups acting
in dimensions $d\geq 3$. (For the case $d=2$ that we have
considered exhaustively any reducible group is diagonal).

Our exhaustive qubit results indicate that the Pauli matrix group
is very special in possessing a suitably rich variety of
normalisers. In this vein it would be particularly interesting to
identify a mathematical ``signature'' property of a given matrix
group $G$ whose validity signals the existence of non-trivial
projective normalisers.

Finally we point out that lemma \ref{clifflemma} asserts a
remarkable structural property of normalisers of the Pauli group
$\cp$ and its tensor powers $\cp^{\otimes n}$ viz. that for levels
$n\geq 3$ there are no new normalisers beyond those generated from
circuits of $n=1$ and 2 normalisers. (This is also true of the
generalised Pauli groups in prime dimension\cite{sean}). The proof
of this property utilises many extra properties special to the
Pauli matrices. Thus we may ask: are there teleportation groups
$G$ (even in dimension $d=2$) such that $G^{\otimes n}$ for $n\geq
3$ has normalisers that are not expressible as composites of
$n=1,2$ normalisers? In our computational analyses we have
considered projective normalisers only up to tensor square
$G\otimes G$ and it remains open whether or not there may exist
entangling normalisers for $n\geq 3$, even in the case that they
are absent for $n=2$. Any mathematical signature property of the
kind mentioned above would be helpful in addressing this
fundamental issue.

\section{Acknowledgements}
We thank Joseph Chuang for informative discussions. SC also thanks
Ashley Montanaro for helpful discussions and acknowledges
financial support from the EPSRC and GCHQ. RJ and NL were
supported in this work by the EPSRC QIP-IRC grant and the EC
project QAP.


\begin{thebibliography}{10}
\bibitem{Gthesis} D.~Gottesman,
\newblock  Stabilizer Codes and Quantum Error Correction,
\newblock PhD thesis, California Institute of Technology, Pasadena, CA, 1997.

\bibitem{NC}  M. A. Nielsen and I. Chuang, Quantum Computation
and Information, CUP 2000.

\bibitem{JL} R. Jozsa and N. Linden, On the role of entanglement in q
uantum-computational speedup, {\em Proc. Roy. Soc. Lond.} A {\bf
459}, 2011-2032 (2003). arXiv:quant-ph/0201143

\bibitem{cliffs} D.~Gottesman,
Course on quantum error correction, Perimeter Institute, Waterloo
http://perimeterinstitute.ca/people/researchers/dgottesman/CO639-2004/index.html

\bibitem{AG} S. Aaronson and D. Gottesman, Improved simulation of
stabiliser circuits, Phys. Rev. A 70:052328, 2004.
quant-ph/0406196.

\bibitem{BK} S. Bravyi and A. Kitaev, Universal quantum
computation with ideal clifford gates and noisy ancillas, Phys.
Rev. A 71, 022316 (2005)

\bibitem{mmtcomp} R.~Raussendorf and H.~J. Briegel,
\newblock A one-way quantum computer.
\newblock {\em Phys. Rev. Lett.}, {\bf 86}, 5188--5191, 2001.
\newblock {arXiv}:quant-ph/0010033;
 R.~Raussendorf, D.~E. Browne, and H.~J. Briegel.
\newblock Measurement-based quantum computation with cluster states
\newblock {\em Phys. Rev. A}, {\bf 68}, 022312, 2003.
\newblock {arXiv}:quant-ph/0301052.

\bibitem{mqc} R. Jozsa, An introduction to measurement based
quantum computation, Proc. NATO-ASI vol. 199, Quantum Information
Processing from theory to experiment,  ed. D Angelakis et al., p.
137-158, 2006.
\bibitem{universal_imprimitive}
J.~{Brylinski} and R.~{Brylinski}.
\newblock {Universal quantum gates}.
\newblock {\em quant-ph/0108062}, August 2001.

\bibitem{blichfeldt}
H. F. Blichfeldt.
\newblock {\em Finite Collineation Groups}.
\newblock University of Chicago press, 1$^{st}$ edition, 1917.

\bibitem{subgroupsSU3}
W.~M. Faibairn, T.~Fulton, and W.~H. Clink.
\newblock Finite and disconnected subgroups of $su_3$ and their spectrum
  application to the elementary-particle spectrum.
\newblock {\em Journal of mathematical physics}, 5(8), August 1964.

\bibitem{subgroupsSU4}
A. Hanany and Y. H. He.
\newblock A monograph on the classification of the discrete subgroups of su(4).
\newblock {\em Journal of High Energy Physics}, 2001(02):027, 2001.

\bibitem{karpilovsky}
G. Karpilovsky.
\newblock {\em Group Representations}.
\newblock North-Holland mathematics studies, $1^{st}$ edition, 1992.

\bibitem{curtisReiner}
C. W. Curtis and I. Reiner.
\newblock {\em Representation theory of finite groups and associative
  algebras}.
\newblock Wiley, $1^{st}$ edition, 1962.

\bibitem{gap}
The~GAP Group.
\newblock Gap --- groups, algorithms, and programming.
\newblock {\em http://www.gap-system.org}, 2005.
\newblock Version 4.4.6.

\bibitem{repsn}
V. Dabbaghian-Abdoly.
\newblock An algorithm to construct representations of finite groups.
\newblock {\em Ph.D. thesis, Dept. Mathematics, Univ. Carleton}, 2003.
\newblock Available as part of GAP software.

\bibitem{wigner}
E. P. Wigner.
\newblock {\em Group theory and its application to the quantum mechanics of
  atomic specra}.
\newblock Academic Press Inc., expanded and improved edition, 1959.

\bibitem{serre}
J.~P. Serre.
\newblock {\em Linear Representations of Finite Groups}.
\newblock Springer-Verlag, $2^{nd}$ edition, 1977.

\bibitem{discreteTorsion}
B. Feng, A. Hanany, Y. H. He, and N. Prezas.
\newblock Discrete torsion, non-abelian orbifolds and the schur multiplier.
\newblock {\em Journal of High Energy Physics}, (01):033, 2001.

\bibitem{MS} W. Malfait and A. Szczepanski, The structure of the
(outer) automorphism group of a Bieberbach group. {\em Composito
Mathematica}, {\bf 136}, p89-101 (2003).

\bibitem{sean} S. Clark, Valence bond solid formalism for d-level
one way quantum computation, J. Phys. A: Math. Gen. 39,
2701-2721(2006). quant-ph/0512155.




\end{thebibliography}
\end{document}